\documentclass[twocolumn,superscriptaddress,aps,floatfix]{revtex4}
\begin{document}
\title{Comment on ``Dynamics of a Charged Particle"  by F. Rohrlich [Phys. Rev. E {\bf 77}, 046609 (2008)]}
\author{N. M. Naumova}
\affiliation{Laboratoire d'Optique Appliqu\'{e}e, UMR 7639 ENSTA, \'Ecole Polytechnique,
CNRS, 91761 Palaiseau, France} 
\author{I. V. Sokolov}
\affiliation{Space Physics Research Laboratory, University of Michigan, Ann
Arbor, MI 48109} 
\begin{abstract}
The equation derived by F. Rohrlich  (Phys. Rev. E {\bf 77}, 046609
(2008)) has been known  
for 60 years (C. J. Eliezer, Proc. Royal Soc. London. Ser. A {\bf 194}, 543 (1948)).
For a long time this equation
has been considered to be incorrect. If there is any need to revisit 
this issue, the only 
relevant consideration is that the Eliezer equation
is very difficult to solve numerically: 
the acceleration 
being expressed in terms of a function 
that, itself, depends
on the acceleration. 
\end{abstract}

\maketitle

The paper \cite{bib:R} claims to derive, after a century of fruitless efforts the following 
equation to describe the motion of an emitting electron (see Eq.(14)):
\begin{equation}
m\dot{v}^\alpha = F^\alpha + \tau_0\dot{F}^\alpha+\tau_0v^\alpha (v_\beta \dot{F}^\beta),\label{eq:1}
\end{equation}
where $m$ is the electron mass, $v^\alpha$ is the electron 4-velocity, $\tau_0=2q^2/(mc^3)\approx 6.2\cdot10^{-24}$ s,
$q$ is the charge of electron, with the choice $c=1$ for the units and $(-+++)$ for a signature. It is claimed that 
the 4-force experienced by the electron, $F^\alpha $, is an arbritrary given function of time. However, while Eq.(\ref{eq:1}) 
being derived, the following restriction on the 4-force has been used:
\begin{equation}
v_\beta F^\beta=0.\label{eq:2} 
\end{equation}
Since this entity should hold for any 4-velocity, the 4-force must 
{\it actually}
be a specific function of 4-velocity
rather than being an arbitrary function of time (for example, for the Lorentz force, $F^\alpha=qF^{\alpha\beta}v_\beta$, the
identity Eq.(\ref{eq:2}) is fulfilled as the result of the anti-symmetry of the field tensor, $F^{\alpha\beta}$). The 
time derivative of the velocity-dependent force, $\dot{F}^\alpha$, must be acceleration-dependent. Moreover, for the 
projection of $F^{\alpha\beta}$ onto the direction of $v^\alpha$ the dependence on the acceleration can be found explicitly. Indeed,
 on differentiating (\ref{eq:2}) one can see, that:
\begin{equation}
v_\beta\dot{F}^\beta = - \dot{v}_\beta F^\beta.\label{eq:3}
\end{equation} 
Eq.(\ref{eq:1}) can be also re-written, using Eq.(\ref{eq:3}):
\begin{equation}
m\dot{v}^\alpha = F^\alpha + \tau_0\dot{F}^\alpha-\tau_0v^\alpha (\dot{v}_\beta F^\beta).\label{eq:4}
\end{equation}
Now we find that:

1. For the case, when the only force experienced by the electron is the
Lorentz force, Eq.(\ref{eq:4}) is well-known 
for 60 (!) years: see Eq.(52) in Eliezer's paper \cite{bib:el}. For a very long time this equation has been considered to be incorrect: 
see the papers reviewed in survey 
\cite{bib:kl}, which is also not new. 
Both Eliezer's paper and the follow-up critical publications are not addressed 
in \cite{bib:R}, although the derived Eq.(14) and the Eliezer equation are entirely identical, 
as demonstrated above. 

2. For the case which is claimed to be considered in \cite{bib:R}, specifically, when 4-force is assumed to be an arbitrary function of 
time, the system Eqs.(\ref{eq:1},\ref{eq:2}) is mathematically
incorrect, being overdetermined. Indeed, Eq.(\ref{eq:3}) is not a consequence of
Eq.(\ref{eq:1}), therefore even if Eq.(\ref{eq:2}) holds in the initial time instant, still the solution of Eq.(\ref{eq:1}) advanced through some 
time interval will break Eq.(\ref{eq:2}), generally
speaking. We emphasize, that the Lorentz force satisfies
Eq.(\ref{eq:2}) by virtue of its {\it explicit} dependence on
4-velocity. 

3.  It is not instructive to revisit the Eliezer equation, but the only new consideration
to be mentioned is that it is difficult to solve 
numerically. Indeed, the acceleration on left hand side of Eq.(\ref{eq:1}) 
is expressed in terms of the right hand side, on which: (1) the last term depends on the acceleration explicitly (see Eq.(\ref{eq:4})) and 
the second term depends on the acceleration implicitly (indeed, nobody knows how).

\end{document}